\documentclass[a4paper]{article}

\usepackage{INTERSPEECH2020}

\title{s-Transformer: Segment-Transformer for Robust Neural Speech Synthesis}
\name{Xi Wang, Huaiping Ming, Lei He, Frank K. Soong}
\address{
  Microsoft (China) Corporation}
\email{\{xwang, huming, helei, frankkps\}@microsoft.com}

\begin{document}

\maketitle
\begin{abstract}
Neural end-to-end text-to-speech (TTS) , which adopts either a recurrent model, e.g. Tacotron, or an attention one, e.g. Transformer, to characterize a speech utterance, has achieved significant improvement of speech synthesis. However, it is still very challenging to deal with different sentence lengths, particularly, for long sentences where sequence model has limitation of the effective context length. We propose a novel segment-Transformer (s-Transformer), which models speech at segment level where recurrence is reused via cached memories for both the encoder and decoder. Long-range contexts can be captured by the extended memory, meanwhile, the encoder-decoder attention on segment which is much easier to handle. In addition, we employ a modified relative positional self attention to generalize sequence length beyond a period possibly unseen in the training data. By comparing the proposed s-Transformer with the standard Transformer, on short sentences, both achieve the same MOS scores of 4.29, which is very close to 4.32 by the recordings; similar scores of 4.22 vs 4.2 on long sentences, and significantly better for extra-long sentences with a gain of 0.2 in MOS. Since the cached memory is updated with time, the s-Transformer generates rather natural and coherent speech for a long period of time. 
\end{abstract}
\noindent\textbf{Index Terms}: speech synthesis, TTS, Transformer, segment recurrence, cached memory

\renewcommand{\thefootnote}{}
\footnotetext{Samples are available at https://vancycici.github.io/sTransformer/}

\section{Introduction}

End-to-end sequence-based TTS models such as Tacotron \cite{Wang2017-Taco} \cite{Shen2018-Taco2} and Transformer TTS \cite{Li2019-Tftts} \cite{Ren2019-Fastspeech} have successfully applied to model \(\langle text,speech\rangle\) pairs. With the help of high quality neural vocoder like WaveNet\cite{Oord2016-Wavenet}, LPCNet \cite{Valin2019-Lpncnet}, GAN based vocoder \cite{Kumar2018-Melgan}, and etc, the generated speech quality could be very natural. As a unified framework, the seq2seq TTS model has largely improved the model capability, taking advantages of direct modeling from text input to speech and whole utterance modeling with either RNN attention or Transformer multi-head attention. 

However, it is still challenging for the seq2seq TTS model to generate extra long sentences, as the ability of modeling complex and long-term dependency remains a research problem. RNNs are difficult to optimize due to gradient vanishing and explosion \cite{Hochreiter2001-Grad} while the sequence length increasing. Transformer \cite{Vaswani2017-Transformer}, based solely on attention mechanisms, with global dependencies between input and output, owing efficient computation and high parallelization, requires large memory growth of sequence length and is difficult to optimize during training on the various length speech corpus, especially when the acoustic feature number arrives to thousands of frames. 

How to capture the long-term dependency has been a generic challenge for the sequence transduction model. Various attention mechanisms are tried in the encoder-decoder attention model, like stepwise monotonic attention \cite{He2019-Step},  forward-attention\cite{Zhang2018-Forward}, monotonic chunkwise attention \cite{Chiu2017-Moca}, and etc. However, these attention mechanisms are usually based on previous state calculation, and not easy to apply into the high parallelized attention calculation of Transformer model. Sparse-Transformer \cite{Child2019-Sparse}, introducing sparse factorization on the attention matrix, gains the ability to produce thousands of timesteps on images, audio and texts, nevertheless  it only works for the self-attention, which is not direct suitable for encoder-decoder attention as different length pair as \(\langle text,speech\rangle\). In the speech recognition tasks, there are many works on the streaming end-to-end model \cite{Jaitly2016-AOS}, using CTC plus Transformer \cite{Moritz2019-TAF}\cite{Moritz2020-SAS} or RNN-Transducer \cite{Alex2012-RNNT} plus LAS \cite{Chan2016-LAS} \cite{Tara2020-ASO}. These two-stage flows need to be jointly decoded or rescored later. Transformer-XL \cite{Dai2019-TXL}, incorporating a segment-level recurrence mechanism and a novel positional encoding schema, achieves excellent performance in language model task for extra long sequence. It reuse the hidden states obtained from previous segments and builds up a recurrent connection between segments. Thus, modeling very long-term dependency becomes possible because information can be propagated through the recurrent connections.

Inspired by the segment streaming model, we think it would be feasible to model speech utterances into segments without disrupting the temporal coherence. We have tried segment based Tacotron model using partial conditioning with incremental prediction\cite{Jaitly2016-AOS}, but due to the size limitation of previous RNN state, the experiment reports inferior performance. Accordingly we turn to Transformer-XL like model which is flexible and high efficient, and propose s-Transformer for speech synthesis. This recurrence mechanism is incorporated in both Transformer encoder and decoder self-attention, as text input and spectrogram output segments, with previous segments computed as hidden states cached as extended context, while the encoder-decoder attention aligns in segment-level. The cached memory lengths for encoder and decoder are setting differently regarding the assumptions that text sequence usually has a long-term dependence while the speech spectrogram is more local related. We set a large cached memory size for encoder, and a smaller one for decoder. With the decoder cached memory, it allows the adjacent acoustic features between segments to participate in the computation as segment overlap, to relax the strict alignment of segment and maintain the coarticulation in continuous speech. As the cached memory size is fixed, the memory will update while segment passing through and “forget” the “too far” segment when it exceeds the memory capacity. What’s more, with the relative positional self attention, it gains generalized attribute regardless of various speech utterance lengths. The s-Transformer can generate extra long utterance which is even unseen in the corpus and obtains the appealing modeling capability beyond sentences like paragraphs.
\begin{figure}[t]
	\centering
	\includegraphics[width=\linewidth]{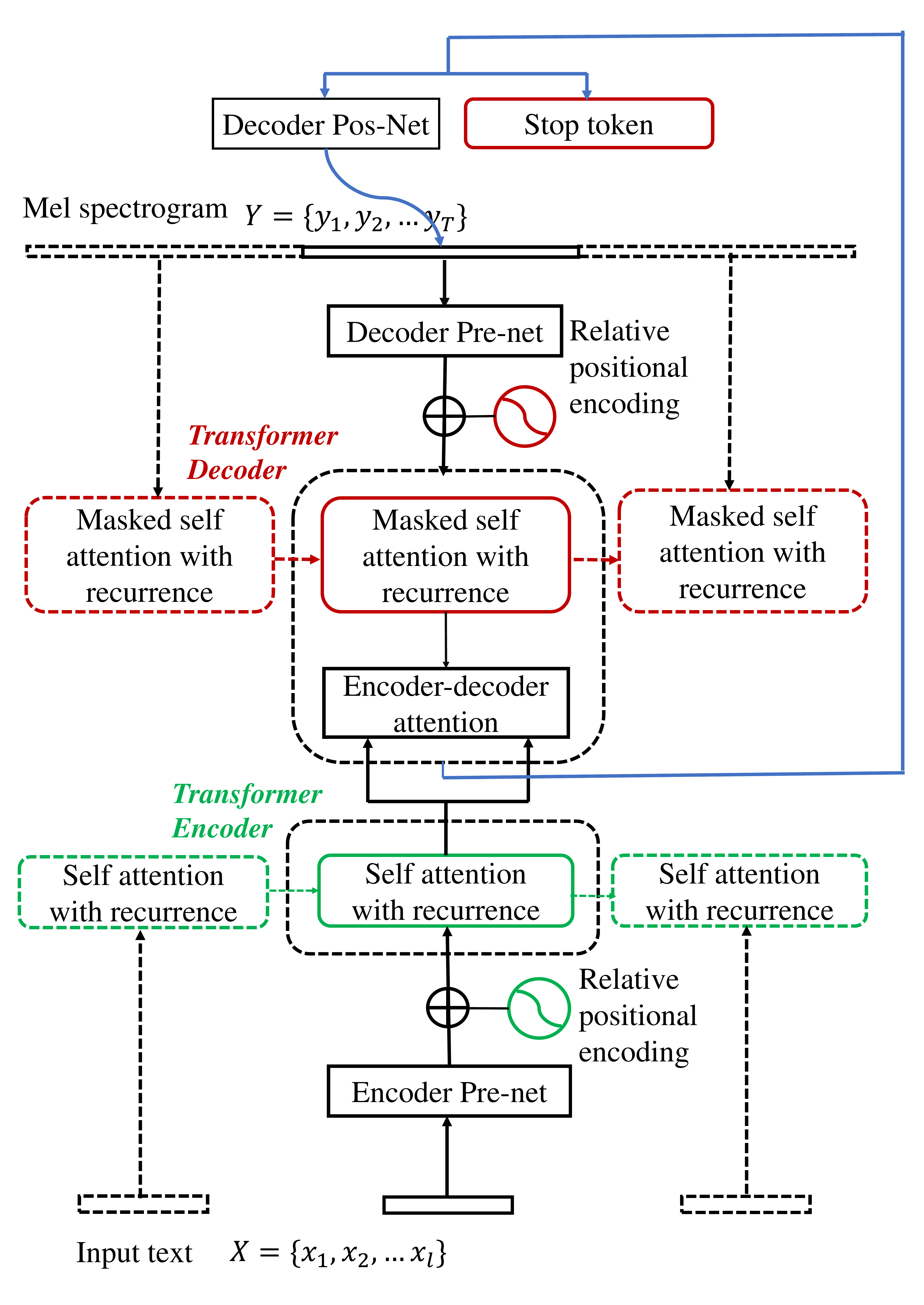}
	\caption{An overview of proposed s-Transformer based speech synthesis. The green and red modules are different as standard Transformer TTS. Dot line modules represent previous and next segments.}
	\label{fig:speech_production}
\end{figure}
\section{s-Transformer}
The proposed s-Transformer based speech synthesis system is shown in Figure 1. Given an input text sequence embedding \(x=\{x_1,x_2,…x_l\}\) and output mel-spectrogram \(y=\{y_1,y_2,…y_T \}\), with alignment, it approximately splits according to a predefined input chunk length as  \(x_\tau=\{x_{\tau,1},x_{\tau,2},…x_{\tau,l} \}\) and corresponding output chunk \(y_\tau=\{y_{\tau,1},y_{\tau,2},…y_{\tau,l^{'}} \}\). Different from Transformer TTS modeling on the whole utterance level, it applies the s-Transformer model on the speech segment, keeping the segments internal order in the utterance.  In the Transformer encoder and decoder blocks, the self-attention modules are calculated with previous segment recurrence as extended context with the help of cached encoder and decoder memories. The Transformer encoder-decoder attention remains the same to produce variable output lengths by using the query vectors and decoder states to generate the attention matrix to a sequence of encoder state. Although the gradients are calculated in the segment, the additional extended context allows the network to exploit the long-range context of previous history information. In other words, the s-Transformer essentially works similar as the monotonic chunk-wise attention with segment recurrence, to ease the optimization for s-Transformer architecture while learning long-term dependency. 

\subsection{Segment level recurrence  }
With the recurrence mechanism to Transformer architecture for consecutive segments, the hidden state sequence computed for the previous segment is cached to be reused as an extended context when the model processes the current segment. Let the two consecutive segments be as \(x_\tau=\{x_{\tau,1},x_{\tau,2},...,x_{\tau,l}\}\) and \(x_{\tau+1}=\{x_{\tau+1,1},x_{\tau+1,2},...,x_{\tau+1,l^{'}}\}\)\textsl{} respectively, denoting the \(\textit{n-th}\) layer hidden state sequence produced for the \(\tau-th \) segment \(s^\tau\). Then, the \(\textit{n-th}\) layer hidden state for segment \(s_{\tau+1}\) is calculated as followings:

\begin{equation}
\tilde{h}=[SG(h^{n-1}_\tau)\circ h^{n-1}_{\tau+1}]
\end{equation}
\begin{equation}
q^{n}_{\tau+1}=h^{n-1}_{\tau+1}\mathbf{W}^\mathrm{T}_{q}
\end{equation}
\begin{equation}
k^{n}_{\tau+1}, v^{n}_{\tau+1}=\tilde{h}^{n-1}_{\tau+1}\mathbf{W}^\mathrm{T}_{k}, \tilde{h}^{n-1}_{\tau+1}\mathbf{W}^\mathrm{T}_{v}
\end{equation}
\begin{equation}
h^{n}_{\tau+1}=\textit{Transformer-laye}r(q^{n}_{\tau+1},k^{n}_{\tau+1},v^{n}_{\tau+1})
\end{equation}

\begin{figure}[t]
	\centering
	\includegraphics[width=\linewidth]{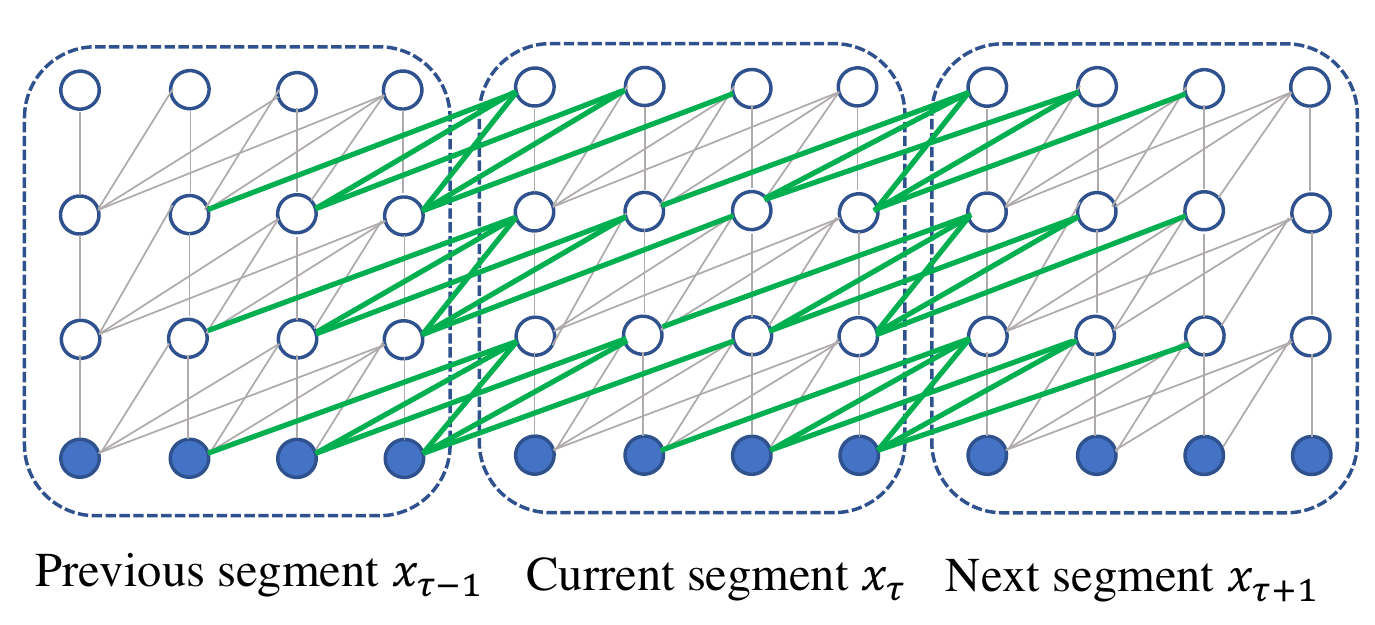}
	\caption{Illustration of segment level recurrence for self attention, with an example of segment length 4.}
	\label{fig:speech_production}
\end{figure}

Where the function \(SG(\cdot)\) stands for stop gradient, the notion \([h_v\circ h_u]\) indicates the concatenation operation for the two sequences along the length dimension as extended context, and \(\mathbf{W}\) denote the Transformer model matrix parameters. The critical difference is the hidden state \(\tilde{h}^{n-1}_{\tau+1}\) which is extended with previous segment. Correspondingly, the current segment based key \(k^{n}_{\tau+1}\) and value \(v^{n}_{\tau+1}\)are derived from this extended hidden state, while query \(q^{n}_{\tau+1}\) remains the same computation from standard hidden state as \(h^{n-1}_{\tau+1}\).
The cached memory could contain multiple previous segments, which extends the context information beyond just two consecutive segments. It’s noticed that the recurrent dependency between connected layers shifts one layer downwards per-segments, which is close to global dependency as Transformer self-attention, using local query to compute the attention with extended hidden state derived key and value. Consequently, the largest dependence length grows linearly with number of layers, segment length and cached memory length. 

\subsection{Relative positional self attention}

In order for the Transformer model to make use of the sequence order, the relative or absolute positional embedding is injected with the input embeddings at the bottoms of the encoder and decoder stacks. In Transformer TTS, we use scaled positional encoding \cite{Li2019-Tftts} instead of fixed positional embeddings, to adaptively fit the scale of both encoder and decoder, regarding the range of text and speech sequence are quite different. The scaled weights fit for both encoder and decoder positional embedding are trainable. 
When we reuse the extended context into hidden state, we should also handle the coherence of position information. In the segment positional embedding, information of relative position in the internal segment sequence is injected by triangle positional embeddings. Meanwhile, similar as  \cite{Shaw2018-SAW}\cite{Huang2018-MTF} introducing relative position representations to allow attention to be informed by how far two positions are apart in the sequence,  we propose to add a trainable bias matrix to learn the relative position embedding for the extended context between a query and key respectively.  
In the standard Transformer TTS, encoder self-attention has a masking bias matrix to mask the padding input embeddings, while decoder self-attention has a casual masking bias matrix to prevent the usage of future mel-spectrograms. Let the masking bias matrix denoted as \(\mathbf{B}^{cur}\) which is based on current segment input embeddings, the trainable bias matrix denoted as \(\mathbf{B}^{rel}\) for the extended context, an \(L\times (L+M)\) dimensional logits matrix multiplied by \(\mathbf{Q}\) and \(\mathbf{K}\), which modulates the attention probabilities for each head as:
\begin{equation}
\textit{RelativeAttention}=\textit{Softmax}(\mathbf{Q}\mathbf(K)^{\mathrm{T}}+ [\mathbf{B}^{rel} \circ \mathbf{B}^{cur}])\mathbf{V}
\end{equation}

\([\mathbf{B}^{rel}\circ \mathbf{B}^{cur}]\) indicates the concatenation of the relative extended bias matrix and corresponding bias matrix along the time length dimension.  

\subsection{Segment and sentence feature}

\subsubsection{Chunk stop token}

Similar as Tacotron2, Transformer-TTS model also employs an utterance stop token to help autoregressive generation stop emitting output while synthesis.  As in the s-Transformer, it should have the capability to know when to stop generating output for this segment. We propose to add a chunk stop token network which works together with utterance stop token. The chunk stop network is a Linear network similar as stop network predicted from decoder output. The final stop logit is weighted summarized by both stop token and chunk stop token jointly to decide when to stop for this segment no matter it’s beginning, middle or ending part in the utterance.
\begin{equation}
\textit{StopLogit}=L^{chunk}(D)\times(1-utt^{end})+L^{utt}(D)\times(1-utt^{end})
\end{equation}
Where \(L^{chunk}\) and \(L^{utt}\) represent the Linear network of chunk stop and utterance stop token, \(D\) denotes the decoder output, and \(utt^{end}\) is the 0/1 indicator as 1 when current segment is the ending of utterances. 

\subsubsection{Chunk speaking rate}
To improve the various speaking rate for multiple speech segments, we propose to add a chunk speaking rate feature into training. The chunk speaking rate is calculated as:
\begin{equation}
\textit{ChunkSpeakingRate}=L^{chunk}/T^{chunk}
\end{equation}
\(L^{chunk}\) and \(T^{chunk}\)represent the chunk input embedding length and corresponding mel-spectrogram frames. 
The chunk speaking rate is calculated as a feature during the training, adding with encoder output. Additionally, there is another chunk speaking rate network as Linear network trained from encoder output. During inference, the chunk speaking rate could be estimated from encoder output and averaged according to the encoder length dimension axis.  

\subsubsection{Sentence feature}
There are some global sentence features, such as the sentence type which is implied in the “future” segment, missing in the s-Transformer while model the current segment. To address this limitation, we collect the sentence feature as additional global embedding input, pass through another Pre-net, add with the segment input embedding as encoder input.  

\section{Experiments}
We apply Transformer TTS as baseline and s-Transformer on a professional enUS female speaker, 45.8 hours, 16kHz, 16bit speech corpus to evaluation the capability on Text-to-Speech voice quality, include short, long and extra long sentences. Besides acoustic model, we train a 20 layer 16kHz 16bit WaveNet neural vocoder conditional by mel-spectrogram to generate the waveform. All the subjective tests are evaluated via Microsoft crowdsourcing UHRS (Universal Human Relevance System) platform, with at least 9 native judges for Comparsion MOS (as CMOS)\cite{Louizou2011-SQA} and 15 judges for MOS test for each test case.  

\subsection{Baseline}
The baseline model have the similar architectures and hyper parameters as Transformer TTS\cite{Li2019-Tftts}. Both encoder and decoder are composed of 6 layers with hidden size as 512 and each multi-head attention has 8 heads, including encoder-decoder attention. We use phoneme as input and 80-channel mel-spectrogram as output, the speech frame shift is 12.5ms. Dynamic batch size is used to support for efficient parallel computing. The model is trained on NVIDIA V100 4xGPU 16GB, with maximum batch size 64 on each GPU, using Adam optimizer \(\beta_{1}=0.9\), \(\beta_{2}=0.999\), \(\epsilon=10^{-8}\) and scheduled learning rate exponentially decay. At first, we remove the long utterances (\(>\)800 frames as 10 seconds, about 20\% in the corpus) to get a stable baseline model, for some long utterances in the corpus will lead the encoder-decoder attention hard to learn. Then, we use all utterances to refine the model. To help the baseline model on extra long sentence, we augment the corpus about 40\% utterances to increase the long sentences percentage by merging two sentences into one, from 1200 to 2400 frames length utterances increase about 20\% for further model refinement. 

\subsection{s-Transformer}
We employ the same parameters as Transformer TTS, with 6 layers for both encoder and decoder with hidden size as 512, self multi-head as 8, but the encoder-decoder multi-head attention number is reduced to 4 considering the segment alignment task would be easier. The encoder and decoder Pre-net and output Pos-net are the same as Baseline. 

\subsubsection{Order chunk reader}
There are some additional parameters for s-Transformer: chunk size for input segment, cached encoder and decoder memory length. The speech corpus is aligned by a speech recognition model on phoneme duration. Chunk size for input segment is calculated for the input phonemes including punctuation, delimiter and etc., and corresponding me-spectrogram segment is obtained according to the phoneme alignment. The chunk size is not exactly segmention point, but to find the center phone. To preserve the speech coarticulation for phonemes in a word, a search window from center phone left to right is set as 20 to find an appropriate segmentation point, preferring punctuation or word boundary. The data reader supports to fetch the segments in utterance order within a fixed batch size.

\begin{table*}[th]
	\caption{CMOS test for different domains. \# calculate the average value. Input number include phoneme, punctuation, delimiter and etc. 800, all and aug. means different training corpus, while aug. represent enlarged augmented extra-long sentences.}
	\label{tab:example}
	\centering
	\begin{tabular}{cccclccc}
		\toprule 
		Domain     & \#input & \#word &\#$\approx$speech   & Model           & Baseline  & Baseline & Baseline \\ 
		           & (num)	 & (num)	  &(frames)	       &(params enc mem \& corpus)        & (\(<\)800)    & (all)    & (aug.)   \\
		\midrule
		Short      & 91      & 11.5   & 330               & s-Transformer(120 \& all)    & 0.033     & -0.126   & -            \\
		Long       & 281     & 31.3   & 950               & & 0.304     & -0.012   & -            \\
		\midrule
		Short      & 91      & 11.5   & 330               & s-Transformer(180 \& all)   & -         & 0.048    & -            \\
		Long       & 281     & 31.3   & 950               & & -         &-0.033    & -            \\
		\midrule
		Extra-long & 426     & 48.5   & 1500              & s-Transformer(180 \& aug.)  & -         & -        & 0.415          \\
		\bottomrule
	\end{tabular}
	
\end{table*}

\subsubsection{Training and synthesis}
s-Transformer performs training iteration on each segment batch with fetched cached encoder and decoder memories. The cached memory will be zero initialized at each utterance beginning. We conduct the s-Transformer training on NVIDIA V100 8xGPU 32GB with fixed batch size 16 for each GPU. There is no need to prune the training corpus for s-Transformer is robust to various speech utterance lengths. We explore the following parameter settings for the experiments: chunk size=60, decoder memory=4, encoder memory lengths as 120 and 180 respectively. During synthesis, the phoneme input is also followed the same strategy for the segmentation. The whole utterance is generated segment by segment and merged in order as one utterance. 

\subsection{Evaluation}
In order to test the general voice quality of TTS model and the capability of long-term dependence, we prepare 3 different domains of test set, each set contains 30 sentences, content from news and audiobooks, but different on the text length (total word number) and rough estimated speech length according to synthesized speech. 

Firstly, we conduct a group of CMOS tests to compare the voice quality between different Baseline and s-Transformer models. According to Table1 CMOS results, we observe that comparing with Baseline(\(<\)800), the s-Transformer with encoder memory size 120 could achieve equal naturalness on short sentences, obvious better voice quality on long sentences, due to Baseline model is trained on a tailored corpus without over 800 frames long utterances. After refinement with all corpus, the Baseline model is improved on both short and long sentences. The s-Transformer with encoder memory size 180 improve the voice quality on both short and long sentence comparing with Baseline(all), as expected that longer context would help to improve the overall prosody and quality. To further test the capability on extra-long sentence and impact of training corpus, we perform model refinement on the augmented corpus included a certain degree of augmented extra-long training utterances for both Baseline and s-Transformer. We still observes many attention errors occurs when the input text is too long for Baseline model, especially the intelligibility drops a lot for Baseline model synthesis beyond 1500 frames, while s-Transfomer performs not much change. After removing some serious attention error samples in the extra-long test set, the CMOS=0.415 indicates that s-Transformer has obvious advantage on the extra-long sentences.

Then, we select the Baseline (all) and s-Transformer with encoder memory 180 (all) for short and long sentences, and Baseline(aug.) and s-Transformer(aug.) for extra-long sentences to perform MOS test, to evaluate the general naturalness with the recording in similar domain. For the short and long domains, Baseline and s-Transformer gain equal 4.29 MOS score, very close to the recording as gap=0.03 on short sentences, bigger gap=0.2 on long sentences. Specifically, for extra-long test set, s-Transformer significant outperform Baseline with MOS score 3.99 versus 3.79. It performs very stable on the extra-long sentences, and in theory, it even be able to generate speech as long as you want. However, due to extra-long sentence like audiobook content is very challenging to obtain rich prosody, s-Transformer still has MOS gap 0.6 with recording, suggesting that even longer-term context by enlarging the cached memory and extracting higher level history and future information are potential beneficial.    

\begin{table}[th]
	\caption{MOS test for different domains}
	\label{tab:example}
	\centering
	\begin{tabular}{cccc}
		\toprule 
		Domain      & Recording    & Baseline  & s-Transformer \\  
		\midrule
		Short       & 4.32         & 4.29      & 4.29          \\
		Long        & 4.46         & 4.22      & 4.2           \\ 
		Extra-long  & 4.63         & 3.79      & 3.99          \\
		\bottomrule
	\end{tabular}
	
\end{table}

At last, we select a extra long sentence sample as case study to observe the encoder-decode attention which represent the alignments for baseline and concatenate segment alignments for s-Transformer. In the Figure 3, comparing with Baseline, in particular the alignment over 1000 decoding steps, s-Transformer alignment is much clearer and smoother, which is essentially like monotonic chunk-wise attention.   

\begin{figure}[th]
	\centering
	\includegraphics[width=\linewidth]{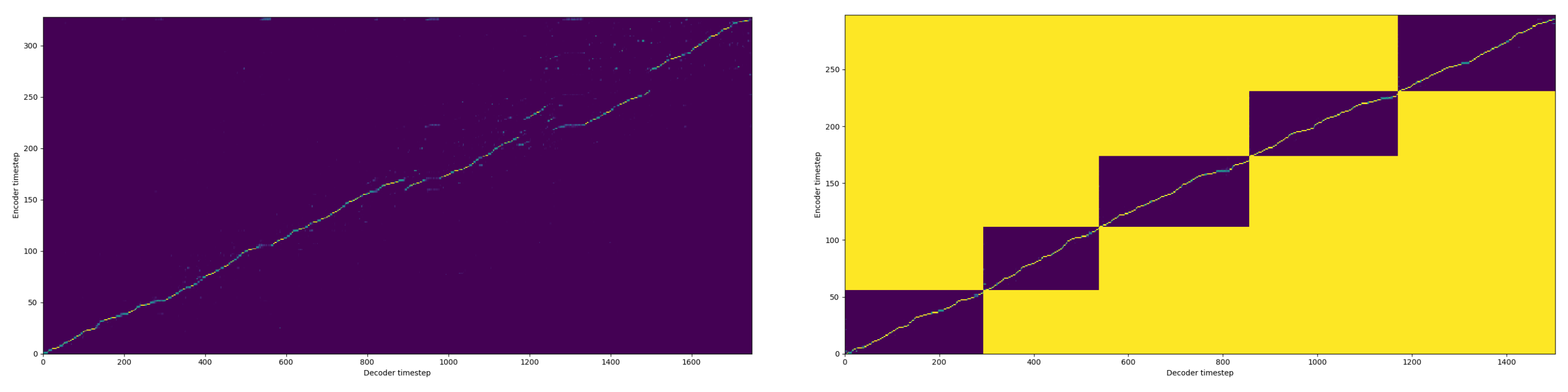}
	\caption{Baseline vs. s-Transformer attention alignment, baseline from layer 0 head 1, s-Transfomer concatenates from 5 segments, layer 1 head 0}
	\label{fig:speech_production}
\end{figure}
 
\section{Conclusions}

In this paper, we propose a novel architecture as s-Transformer model for speech synthesis, which utilizes segment recurrence to capture long-term dependence and performs segment-level encoder-decoder attention to handle the difficulty of long pair \(\langle text,speech\rangle\). Our experiments show the proposed s-Transformer can achieve equal naturalness as standard Transformer TTS on general test set, in particular stable and obvious better quality on extra-long sentences, even unseen long speech as couple minutes as thousands frames. With this segment recurrence mechanism, it builds up a framework for long speech synthesis even beyond sentences.

\clearpage

\bibliographystyle{IEEEtran}

\bibliography{mybib}

\begin{thebibliography}{10}
\providecommand{\url}[1]{#1}
\csname url@samestyle\endcsname
\providecommand{\newblock}{\relax}
\providecommand{\bibinfo}[2]{#2}
\providecommand{\BIBentrySTDinterwordspacing}{\spaceskip=0pt\relax}
\providecommand{\BIBentryALTinterwordstretchfactor}{4}
\providecommand{\BIBentryALTinterwordspacing}{\spaceskip=\fontdimen2\font plus
\BIBentryALTinterwordstretchfactor\fontdimen3\font minus
  \fontdimen4\font\relax}
\providecommand{\BIBforeignlanguage}[2]{{%
\expandafter\ifx\csname l@#1\endcsname\relax
\typeout{** WARNING: IEEEtran.bst: No hyphenation pattern has been}%
\typeout{** loaded for the language `#1'. Using the pattern for}%
\typeout{** the default language instead.}%
\else
\language=\csname l@#1\endcsname
\fi
#2}}
\providecommand{\BIBdecl}{\relax}
\BIBdecl

\bibitem{Wang2017-Taco}
Y.~Wang, R.~Skerry-Ryan, D.~Stanton, Y.~Wu, R.~J. Weiss, N.~Jaitly, Z.~Yang,
  Y.~Xiao, Z.~Chen, and S.~B. et~al., ``Tacotron: Towards end-to-end speech
  synthesis,'' in \emph{Proc Interspeech}, 2017, pp. 4006--4010.

\bibitem{Shen2018-Taco2}
J.~Shen, R.~Pang, R.~J. Weiss, M.~Schuster, N.~Jaitly, Z.~Yang, Z.~Chen,
  Y.~Zhang, Y.~Wang, and R.~S.-R. et~al., ``Natural tts synhtesis by
  conditioning wavenet on mel spectrogram predictions,'' in \emph{IEEE
  International Conference on Acoustics, Speech and Signal Processing(ICASSP)},
  2018, pp. 4779--4783.

\bibitem{Li2019-Tftts}
N.~Li, S.~Liu, Y.~Liu, S.~Zhao, M.~Liu, and M.~Zhou, ``Neural speech synthesis
  with transformer network,'' in \emph{AAAI Conference on Artificial
  Intelligence (AAAI)}, 2019.

\bibitem{Ren2019-Fastspeech}
Y.~Ren, Y.~Ruan, X.~Tan, Q.~Tao, S.~Zhao, Z.~Zhao, and T.~Liu, ``Fastspeech:
  Fast, robust and controllable text to speech,'' in \emph{Neural Information
  Processing Systems}, 2019, pp. 3165--4174.

\bibitem{Oord2016-Wavenet}
A.~van~den Oord, S.~Dieleman, H.~Zen, K.~Simonyan, O.~Vinyals, A.~Graves,
  N.~Kalchbrenner, A.~Senior, and K.~Kavukcuoglu, ``Wavenet: A generative model
  for raw audio,'' in \emph{ArXiv preprint arXiv:1609.03499}, 2016.

\bibitem{Valin2019-Lpncnet}
J.~Valin and J.~Skoglund, ``Lpcnet: Improving neural speech synthesis through
  linear prediction,'' in \emph{Proc. ICASSP 2019, Brighton, United Kingdom},
  2019, pp. 5891--5895.

\bibitem{Kumar2018-Melgan}
K.~Kumar, R.~Kumar, T.~de~Boissiere, L.~Gestin, W.~Z. Teoh, A.~d.~B. J.~Sotelo,
  Y.~Bengio, and A.~Courville, ``Melgan: Generative adversarial networks for
  conditional waveform synthesis,'' in \emph{Neural Information Processing
  Systems}, 2019, pp. 5891--5895.

\bibitem{Hochreiter2001-Grad}
S.~Hochreiter, Y.~Bengio, P.~Frasconi, and J.~Schmidhuber, ``Gradient flow in
  recurrent nets: the difficulty of learning long-term dependencies,'' in
  \emph{Advances in Neural Information Processing Systems}, 2001.

\bibitem{Vaswani2017-Transformer}
A.~Vaswani, N.~Shazeer, N.~Parmar, J.~Uszkoreit, L.~Jones, A.~N. Gomez,
  L.~Kaiser, and I.~Polosukhin, ``Attention is all you need,'' in \emph{Neural
  Information Processing Systems}, 2017, pp. 6000--6010.

\bibitem{He2019-Step}
M.~He, Y.~Deng, and L.~He, ``Robust sequence-to-sequence acoustic modeling with
  stepwise monotonic attention for neural tts,'' in \emph{Proc. Interspeech},
  2019, pp. 1293--1297.

\bibitem{Zhang2018-Forward}
J.-X. Zhang, Z.-H. Ling, and L.-R. Dai, ``Forward attention in
  sequenceto-sequence acoustic modeling for speech synthesis,'' in \emph{IEEE
  International Conference on Acoustics, Speech and Signal Processing(ICASSP)},
  2018, pp. 4789--4793.

\bibitem{Chiu2017-Moca}
C.-C. Chiu and C.~Raffel, ``Monotonic chunkwise attention,'' in \emph{arXiv
  preprint arXiv:1712.05382}, 2017.

\bibitem{Child2019-Sparse}
R.~Child, S.~Gray, A.~Radford, and I.~Sutskever, ``Generating long sequences
  with sparse transformers,'' in \emph{arXiv preprint arXiv:1904.10509}, 2019.

\bibitem{Jaitly2016-AOS}
N.~Jaitly, D.~Sussillo, Q.~V. Le, O.~Vinyals, I.~Sutskever, and S.~Bengio, ``An
  online sequence-to-sequence model using partial conditioning,'' in
  \emph{Neural Information Processing Systems}, 2016.

\bibitem{Moritz2019-TAF}
N.~Moritz, T.~Hori, and J.~L. Roux, ``Triggered attention for end-to-end speech
  recognition,'' in \emph{Proceedings of ICASSP. IEEE}, 2019, pp. 5666--5670.

\bibitem{Moritz2020-SAS}
------, ``Streaming automatic speech recognition with the transformer model,''
  in \emph{Proceedings of ICASSP. IEEE}, 2020.

\bibitem{Alex2012-RNNT}
A.~Graves, ``Sequence transduction with recurrent neural networks,'' in
  \emph{arXiv preprint arXiv:1211.3711}, 2012.

\bibitem{Chan2016-LAS}
W.~Chan, N.~Jaitly, Q.~Le, and O.~Vinyals, ``Listen, attend and spell: A neural
  network for large vocabulary conversational speech recognition,'' in
  \emph{Proceedings of ICASSP. IEEE}, 2016, pp. 4960--4964.

\bibitem{Tara2020-ASO}
T.~N. Sainath, Y.~He, B.~Li, A.~Narayanan, R.~Pang, A.~Bruguier, S.~Chang, and
  et~al., ``A streaming on-device end-to-end model surpassing server-side
  conventional model quality and latency,'' in \emph{Proceedings of ICASSP.
  IEEE}, 2020.

\bibitem{Dai2019-TXL}
Z.~Dai, Z.~Yang, Y.~Yang, W.~W. Cohen, J.~Carbonell, Q.~V. Le, and
  R.~Salakhutdinov, ``Transformer-xl: Attentive language models beyond a
  fixed-length context,,'' in \emph{arXiv preprint arXiv:1901.02860}, 2019.

\bibitem{Shaw2018-SAW}
P.~Shaw, J.~Uszkoreit, and A.~Vaswani, ``Self-attention with relative position
  representations,'' in \emph{arXiv preprint arXiv:1803.02155}, 2018.

\bibitem{Huang2018-MTF}
C.-Z.~A. Huang, A.~Vaswani, J.~Uszkoreit, I.~Simon, C.~Hawthorne, N.~Shazeer,
  A.~M. Dai, M.~D. Hoffman, M.~Dinculescu, and D.~Eck, ``Music transformer:
  Generating music with long-term structure,'' in \emph{Seventh International
  Conference on Learning Representations}, 2018.

\bibitem{Louizou2011-SQA}
P.~C. Loizou, ``Speech quality assessment,'' \emph{Multimedia analysis,
  processing and communications}, pp. 623--654, 2011.

\end{thebibliography}


\end{document}